\documentclass[a4paper]{article}

\usepackage{blindtext}
\usepackage{multicol}
\usepackage{amsmath}
\usepackage{graphicx}
\usepackage{authblk}
\usepackage{hyperref}
\usepackage{indentfirst}
\usepackage[labelfont=bf]{caption}
\usepackage{float}

\usepackage{geometry}
\usepackage[T2A]{fontenc}
\usepackage[utf8]{inputenc}
\usepackage[english]{babel}

\begin{document}

\title{Interpretation of the spectral inhomogeneity in the 10TV region in terms of a close source}
\author[1]{I.Kudryashov}
\author[1]{F.Gasratov}
\author[2]{V.Yurovskiy}
\author[3]{V.Latonov}
\affil[1]{SINP MSU, Moscow, Russia}
\affil[2]{Department of Physics, MSU, Moscow, Russia}
\affil[3]{Department of Mechanics and Mathematics, MSU, Moscow, Russia}
\affil[ ] {e-mails: \href{mailto:ilya.kudryashov.85@gmail.com}{ilya.kudryashov.85@gmail.com}, \href{mailto:vova@txlib.ru}{vova@txlib.ru} }

\date{}

\maketitle

\begin{abstract}

The description of the inhomogeneity of the cosmic ray spectrum in the region of 10 TV, which is observed in experimental data, in terms of isotropic diffusion from a single close source is considered. It is shown that such a description is possible, the area of possible localization of the source in space and time, its energy is found. The method of penalty functions is used to account for the data on the spectrum of all particles.
\end{abstract}

\newpage

\section{Introduction}

Some cosmic rays physics experiments (CR) [1–3] finds evidences of variation of CR spectrum index in area of magnetic rigidity about 10 TV. Let us call it small knee. The data of direct space experiment NUCLEON [1] allow us to resolve single-particle structure of small knee for each abundant primary CR component and shows, that this bending occurs near the same magnetic rigidity 10 TV for each nucleus regardless of Z.

Such an inhomogeneity in the regular CR spectrum can be explained by several reasons: CR production mechanisms (for example, the acceleration limit in the envelopes of a certain type of supernova), propagation mechanisms, or the contribution of a single close source to the CR flux [4]. The significant sharpness of the bending in terms of the magnetic rigidity spectrum [1, 3] is an indirect evidence that the small knee is determined by the CR acceleration limit in a single close source such as a supernova remnant. If this were the contribution of several sources, then it would be difficult to expect such a sharp bending. Therefore, in this article we study the possibility of explaining the slight inflection of cosmic rays spectrum by contribution of one close source.

To test this hypothesis, a mathematical model was developed for describing the spectral features of cosmic rays in terms of diffusion from a close source, first proposed by Erlykin and Wolfendale [5].

The authors have carried out an approximation of the free parameters of the model based on experimental data, constructed domains of parameters for an admissible and most probable single source. A significant difference from previous works of this type, related to the large knee of Kulikov – Christiansen cosmic rays near 3 PeV in energy per particle, is the observation of the small knee not only in the total spectrum of all particles, but also in the spectra of individual abundant nuclei. That is, we are dealing with much more detailed information than is currently available for the 3 PeV CR knee.

\section{Calculation model}

The mathematical model of the expected flux was constructed as the sum of the contribution of the close source and the galactic CR background (Fig. 1):

\[
F_{sum}(R) = F_{bgr}(R) + F_{star}(R)
\]

The power-law spectrum was chosen as the galactic background $F = a_0$ and $\gamma$, where parameters $a_0$ and $\gamma$ for each CR nucleus correspond to the slopes of the spectra, measured in area 50 GV – 3000 GV by experiment data of ATIC [6], NUCLEON [1], AMS-2 [7], CREAM[8] and TRACER [9].

\begin{figure}[h]
    \centering
    \includegraphics[width=0.7\textwidth]{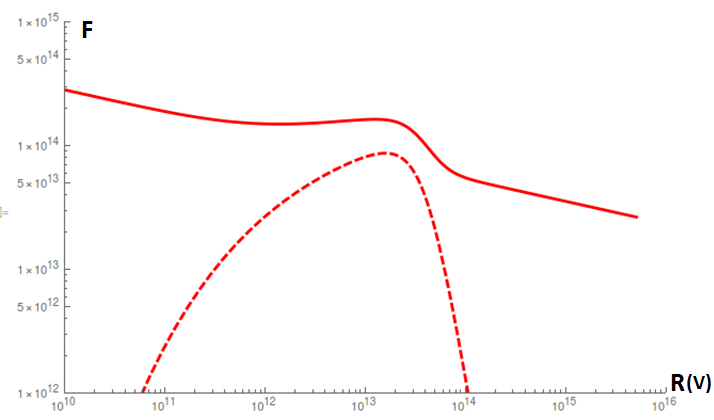}
    \caption{Model signal + background in the proximity of the flash source: the dotted line is the signal from the source, the solid line is the signal together with the background.}
    \label{fig:fig1}
\end{figure}

The contribution of a close source is calculated by solving the diffusion equation in the flash source approximation (instantaneous in time and point in space). This approximation describes well the spatial localization of sources such as supernova remnants, since the distances to such sources are much larger than their sizes. The approximation works well for relatively old sources of cosmic rays (ten thousand years or more), but can only give a qualitatively correct picture for younger supernova remnants.

The spectrum in the source is specified by a double power law with a break and smooth matching of two power-law spectra at the break point:

\[
Q(R, t, r) = R^{-\gamma} (1 + (\frac{R}{R_{ref}})^{\omega_0})^{-\frac{\delta\gamma}{\omega_0}}\delta(r)\delta(t),
\]

where $R$ is the magnetic rigidity, $\gamma$ is the spectrum index before the bending, $\delta\gamma$ is the difference in the spectral indices before and after the bending (moreover, $\delta\gamma$ is a free parameter), $\omega_0$ is the bending smoothing coefficient, $R_ref$ is the position of the bending with $\gamma = 2$. It is assumed that the shape of the spectra is the same for all CR components, and only the integral intensity is different for different CR components.

The diffusion equation for a close source in the isotropic approximation has the form:

\begin{equation}
\frac{\partial N}{\partial t} - \nabla (D \nabla N) = Q(R, t, r), 
\end{equation}

where $N$ is the CR concentration, $Q$ is the source function and $D$ is the diffusion coefficient calculated by the formula:

\[
D_{xx}(R) = D_{xx0}(\frac{R}{R_0})^\delta,
\]

where $D_{xx0} = 4.3 \cdot 10^{28} cm^2s^{-1}$, $\delta = 0.395$ and $R_0 = 4.5 GV$ are parameters taken from work [10].

Since the flux of an instantaneous point source with a certain magnetic rigidity is simply expressed by the Green's function of the diffusion equation, the flux of cosmic rays F, which satisfies Eq. (1) for a point source in the approximation of an instantaneous flash with a spectrum Q (R), is calculated as:

\[
F(R, t, r) = \frac{c}{4\pi}G(R, t, r) Q(R)
\]

where $G(R, t, r)$ is the Green's function for three-dimensional diffusion in infinite space.

At this stage, we do not discuss the contribution of a close source to the anisotropy of GCR, since this is a rather difficult issue that requires special analysis. This contribution will depend on the superposition of the source position and the direction of the local interstellar magnetic field (since the local diffusion tensor is significantly anisotropic ($D_\parallel/D_\bot > 10$) [11] and can vary widely [12, 13]). The model of diffusion transport, taking into account the change in the ratio of the components of the tensor of local ($D_\parallel/D_\bot > 10$) and global diffusion ($D_\parallel/D_\bot ~ 2$) [14], will be considered in the future.

\begin{equation}
\boldsymbol b = \sum_n{A_n \boldsymbol P_n cos(\boldsymbol r \boldsymbol k_n + \phi_n) }
\end{equation}

\section {Experimental data approximation}

The authors use single-particle data from direct CR experiments available at the current time in the rigidity range from 100 GV to 100 TV: for protons and helium, the experiments are NUCLEON [1], AMS-02 [7], ATIC [6], CREAM [8], PAMELA [15], CALET [16] and DAMPE [17], for C and O - NUCLEON, AMS-02, ATIC, CREAM.

The authors use single-particle data from direct CR experiments available at the current time in the rigidity range from 100 GV to 100 TV: for protons and helium, the experiments are NUCLEON [1], AMS-02 [7], ATIC [6], CREAM [8], PAMELA [15], CALET [16] and DAMPE [17], for C and O - NUCLEON, AMS-02, ATIC, CREAM.

Thus, summing the spectra of all abundant nuclei allowed us to construct a model spectrum of all particles, which allowed us to take into account the data of the indirect HAWC experiment [18].

To search for a hypothetical close source describing the discussed spectrum inhomogeneity, it is necessary to optimize its parameters in terms of position in space, age, explosion energy (assuming that the CR source is a supernova remnant). To find the optimal age and distance to the source {t, r}, the functional was minimized:

\[
\chi^2 = \sum_i(\frac{f_i^{mod}-F_i}{\sigma_i})^2,
\]

where $f_i^{mod}$ - simulated particle flows, $F_i$ - experimental flows, $\sigma_i$ - the corresponding experimental error; the summation is carried out over all available experimental points i of different Z-spectra of various experiments. In other words, optimization is carried out for all available direct experiments at once, taking into account the availability of information on individual cores.

Functional also allows to specify the position of the knee in terms of rigidity, change of the spectral slope, power of source, the values of penalty parameters, the concentration of nuclei of helium, carbon and oxygen (normalized to hydrogen prevalence). The penalty parameters will be discussed later.

\section{Penalty method}

The indirect HAWC experiment [18] provides data of the spectrum of all particles. However, the experimental points have large systematic uncertainties: they greatly exceed the values of statistical uncertainties. In order for the data of this experiment to correlate with the data of other experiments, we decided to apply the penalty method (two-dimensional case). It is based on taking into account the limitations imposed on the arguments of the function by transforming the objective function of the original optimization task. In particular, the method was used in quantum chromodynamics [19].

In our case, the objective function is $\chi^2$. A correlation function of the form $\mu = aE + b$ is introduced, where E is the energy, a and b are the penalty parameters. Parameter a is associated with the tilt, parameter b is associated with the offset along the ordinate axis. Taking into account the correlation function, the functional $\chi^2$ looks like this [20]:
\[
\chi^2(\xi, \alpha) = \sum_i \frac{(F_i(1+\sum_j\frac{\partial \mu_i}{\partial \alpha_j}\Delta\alpha_j) - P_i(\xi))^2}{\delta_i^2(1+\sum_j\frac{\partial \mu_i}{\partial \alpha_j}\Delta\alpha_j} + \sum_i \sum_j \Delta \alpha_i \Delta \alpha_j (A_s)_{ij}^{-1},
\]

where $\xi$ are the arguments of the simulated flows $P_i(\xi)$, $\alpha$ is a set of penalty parameters (in our case there are two of them), $F_i$ is experimental flows, $\delta_i$ is experimental uncertainties, As is the correlation matrix. The matrix itself looks like this:
\[
A_s = 
\begin{bmatrix}
\sigma_1^2 & \rho \sigma_1 \sigma_2 \\
\rho \sigma_1 \sigma_2 & \sigma_2^2
\end{bmatrix},
\sigma_1 = \frac{B}{AB-C^2},
\sigma_2 = \frac{A}{AB-C^2},
\rho = \frac{-C}{AB}.
\]

The coefficients $A$, $B$ and $C$ depend on the energy $E_i$ (except $B$) and experimental uncertainties normalized to the flow $\sigma_i^{rel}$:

\[
A = \sum_i^n \frac{E_i^2}{{\sigma_i^{rel}}^2}, 
B = \sum_i^n \frac{1}{{\sigma_i^{rel}}^2},
C = \sum_i^n \frac{E_i}{{\sigma_i^{rel}}^2}
\]

\section{Results and discussion}

For each point in the space {t, r}, the value of $\xi^2$ was minimized according to the parameters of the energy and chemical composition of the source. The resulting surface $\xi^2 (t, r)$ has a complicated shape with a pronounced minimum region.

The region of space {t, r} with the required energy of the source $W < 10^{51}$ erg is considered acceptable, since the power of the supernova explosion according to modern concepts [21] does not exceed this value. The calculations assumed that one tenth of the total energy of the explosion goes into the energy of cosmic rays.

Figure 2 shows a map of the lines of the level $\xi^2/NDF$ (by one degree of freedom): the darker the shade, the less $\xi^2$. The age of the source in thousands of years is plotted along the abscissa axis, and the distance to the source in the kpc is plotted along the ordinate axis. The red winding line corresponds to the energy level $W = 10^{50}$ erg, the green one corresponds to the level $W = 10^{51}$ erg.

Figure 2 shows that the position in the space {t, r} of the hypothetical source is localized in a narrow region, which is an extended region of the minimum on the surface relief. The optimal position of the source in the space {t, r} corresponds to the region of 0.1-0.2 kpc and the age from 1 to 5 thousand years, although the hypothetical source can be located anywhere in the extended minimum region.  The energy of the source is between $10^{50}$ and $10^{51}$ erg. The expected spectra corresponding to the source of 3 thousand years, 0.16 kpc, togeth er with some experimental data used in the analysis, are shown in Fig. 3,4.

\section{Conclusion}

A model of the contribution of a single point source-flash to the background spectrum of CL in the approximation of diffusion without energy losses and fragmentation is proposed to explain the nature of the observed spectral inhomogeneity of CR. The model demonstrates reasonable agreement with experimental data at the source energy up to $10^{51}$ erg, localizes the position of a hypothetical source in the distance-time space in a narrow region of phase space, and also predicts the most likely area of existence of such a hypothetical source at 0.1 – 0.2 kpc and an age of 1 to 5 thousand years.  It should be noted that the optimal source is obtained quite young, so the approximation of the source-flash for its description is not very accurate, and given that the isotropic diffusion for these energies is a very rough approximation, the presented results should be considered preliminary, and in the subsequent work we assume to take into account the anisotropy of the diffusion tensor of the evolution of the supernova remnant at the Sedov-Taylor stage. Thus, it is demonstrated that the explanation of the observed spectral inhomogeneity of the CR near 10 TV in terms of magnetic rigidity by the contribution of a single remnant of a close supernova to the observed cosmic ray fluxes is possible.

\begin{figure}[H]
    \centering
    \includegraphics[width=0.5\textwidth]{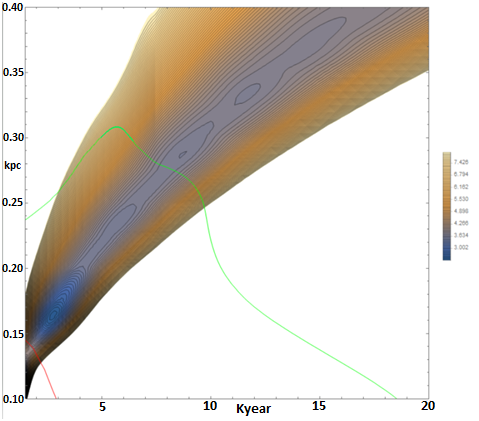}
    \caption{The map of the lines of the level of the function $\xi^2$ by one degree of freedom. The value $\xi^2$ is also transmitted in the color scale.}
    \label{fig:fig2}
\end{figure}

\begin{figure}[H]
    \centering
    \includegraphics[width=0.5\textwidth]{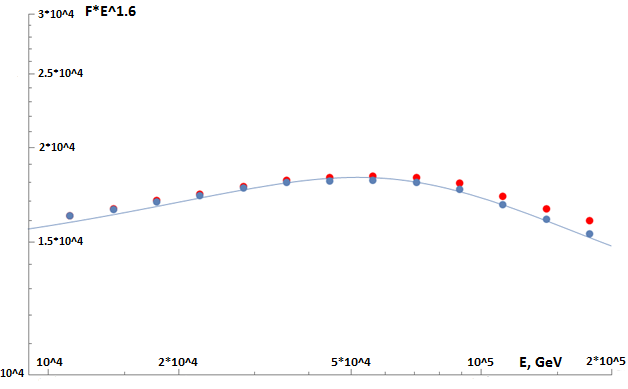}
    \caption{The spectrum of all particles (solid line) together with the data of the HAWC experiment: the starting points are marked in red, the points taking into account the penalty parameters are blue.}
    \label{fig:fig4}
\end{figure}

\begin{figure}[H]
    \centering
    \includegraphics[width=0.5\textwidth]{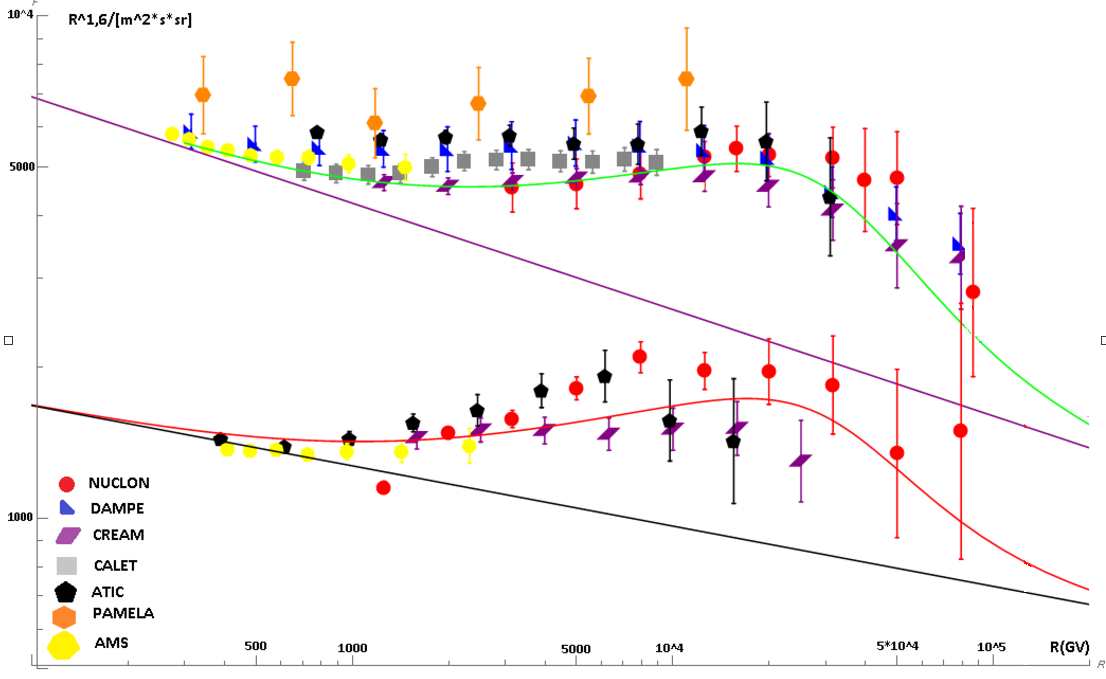}
    \caption{The spectra of protons and helium obtained in the model for a source with an age of 3        thousand years, at a distance of 0.16 kpc, together with the experimental data used for approximation.}
    \label{fig:fig3}
\end{figure}

\newpage

\section*{References}

\renewcommand{\labelenumi}{[\arabic{enumi}]}

\begin{enumerate}
    \item Atkin E., Bulatov V., Dorokhov V. et al. // JETP Lett. 2018. V. 108. No. 1. P. 513.
    \item Atkin E., Bulatov V., Dorokhov V. et al. // JETP Lett. 2018. V. 108. No. 1. P. 513.
    \item Yoon Y.S., Anderson T., Barrau A. et al. // arXiv: 1704.02512. 2017.
    \item Guo Y.Q., Yuan Q. // Chin. Phys. C. 2018. V. 42. No. 7. Art. No. 075103.
    \item Erlykin A.D., Wolfendale A.W. // J. Phys. G. 1997. V. 23.  P. 9.
    \item A. D. Panov, J. H. Adams, H. S. Ahn, et al..// Bulletin of the Russian Academy of Sciences, Physics, 73:564–567, June 2009.
    \item Aguilar, M. et al,//Physical Review Letters vol.114, no. 17,2015. doi:10.1103/PhysRevLett.114.171103.
    \item Y. S. Yoon, T. Anderson, A. Barrau, Net al. // Astrophys. J., 839:5, April 2017.
    \item P. J. Boyle, for the TRACER project.  //Advances in space research Volume 42, Issue 3,5 August 2008, Pages 409-416.
    \item Boschini M.J., Della Torre S., Gervasi M. et al. // Astrophys. J. 2017. V. 840. No. 2. P. 115.
    \item Giacinti G., Kachelries M., Semikoz D.V. // J. Cosmol. Astropart. Phys. 2018. V. 2018. No. 7. Art. No. 051.
    \item Becker Tjus J., Merten L. // Phys. Rep. 2020. V. 872. P. 1.
    \item Casse F., Lemoine M., Pelletier G. // Phys. Rev. D. 2002. V. 65. No. 2. Art. No. 023002.
    \item V. L. Ginzburg. Astrophysics of cosmic rays. Moscow: Nauka, 1990.
    \item V. L. Ginzburg. Astrophysics of cosmic rays. Moscow: Nauka, 1990. 
    \item Adriani O. et al. (CALET Collaboration) // Phys. Rev. Lett. 2019. V. 122. No. 18. Art. No. 181102.
    \item DAMPE Collaboration // Nature. 2017. V. 552. No. 7683. P. 63.
    \item R. Alfaro et al. (HAWC Collaboration). // Phys. Rev. D 96, 122001, 2017.
    \item Zhiqing Zhang. // Proceedings for EPS-HEP2013
    \item Roger Barlow. // arXiv:1701.03701v2
    \item B. S. Ishkhanov, I. M. Kapitonov, I. A. Tutyn.//Moscow: URSS, 2019
\end{enumerate}

\end{document}